\documentclass[%
 preprint,
superscriptaddress,
 amsmath,amssymb,
 aps,
]{revtex4-2}

\usepackage[bookmarks,colorlinks]{hyperref}

\usepackage{graphicx}
\usepackage{dcolumn}
\usepackage{bm}

\begin{document}

\clearpage

\centerline{\bf \large Covert Scattering Control in Metamaterials  }
\centerline{\bf \large with Non-Locally Encoded Hidden Symmetry  }
\bigskip

\centerline{Jérôme Sol\textsuperscript{1}, Malte Röntgen\textsuperscript{2}, and Philipp del Hougne\textsuperscript{3,*}}

\bigskip

\centerline{\textsuperscript{1} \small{\textit{INSA Rennes, CNRS, IETR-UMR 6164, F-35000 Rennes, France}}}
\centerline{\textsuperscript{2} \small{\textit{Laboratoire d’Acoustique de l’Université du Mans, UMR 6613, CNRS, F-72085 Le Mans, France}}}
\centerline{\textsuperscript{3} \small{\textit{Univ Rennes, CNRS, IETR-UMR 6164, F-35000 Rennes, France}}}
\centerline{\textsuperscript{*} \small{Correspondence to \href{mailto:philipp.del-hougne@univ-rennes1.fr}{philipp.del-hougne@univ-rennes1.fr}}}

\bigskip
\bigskip
\bigskip
\bigskip

Symmetries and tunability are of fundamental importance in wave scattering control, but symmetries are often obvious upon visual inspection which constitutes a significant vulnerability of metamaterial wave devices to reverse-engineering risks. 
Here, we theoretically and experimentally show that it is sufficient to have a symmetry in the reduced basis of the ``primary meta-atoms’’ that are directly connected to the outside world; meanwhile, a suitable topology of non-local interactions between them, mediated by the internal ``secondary’’ meta-atoms, can hide the symmetry from sight in the canonical basis. 
We experimentally demonstrate \textit{covert} symmetry-based scattering control in a cable-network metamaterial featuring a hidden parity ($\mathcal{P}$) symmetry in combination with hidden-$\mathcal{P}$-symmetry-preserving and hidden-$\mathcal{P}$-symmetry-breaking tuning mechanisms.
First, we achieve physical-layer security in wired communications, using the domain-wise hidden $\mathcal{P}$-symmetry as shared secret between the sender and the legitimate receiver. 
Then, within the approximation of negligible absorption, we report the first tuning of a complex scattering metamaterial \textit{without} mirror symmetry to feature exceptional points (EPs) of $\mathcal{PT}$-symmetric reflectionless states, as well as quasi-bound states in the continuum. 
Finally, we show that these results can be reproduced in metamaterials involving non-reciprocal interactions between meta-atoms, including the first observation of reflectionless EPs in a non-reciprocal system.

\bigskip


\clearpage
\section{Introduction}

The scattering characteristics of a metamaterial are defined by the properties of its constituent meta-atoms as well as how the latter are coupled with each other and how they are coupled to the asymptotic scattering channels. Often, only a subset of all meta-atoms is directly coupled to the asymptotic scattering channels. These ``primary'' meta-atoms can be conceptually distinguished from the remaining ``secondary'' meta-atoms, since the latter can be equivalently treated as additional non-local interactions between the primary meta-atoms. An important design consideration for a metamaterial, in addition to its ultimate scattering properties and hence functionalities, may be the extent to which the latter can be deduced upon visual inspection of its meta-atoms in the canonical representation, for instance, if reverse engineering is a concern. Of particular vulnerability are devices based on parity ($\mathcal{P}$) and related symmetries. A solution may lie in the degrees of freedom offered by the secondary meta-atoms: whereas a representation of the metamaterial reduced to the primary meta-atoms would reveal the $\mathcal{P}$ symmetry, the $\mathcal{P}$ symmetry can be (although usually it is not) absent in the canonical representation. In this Article, we experimentally demonstrate covert scattering control in cable-network metamaterials featuring (i) a hidden $\mathcal{P}$ symmetry covertly encoded into the topology of the secondary meta-atoms, as well as (ii) hidden-$\mathcal{P}$-symmetry-preserving (and hidden-$\mathcal{P}$-symmetry-breaking) tunability. Specifically, we leverage the absence of mirror symmetries to present a scheme for physically secure wired communications and we observe reflectionless exceptional points (EPs) despite the absence of simple mirror symmetries. Moreover, we show that such scattering control is also feasible in the case of non-reciprocal interactions between the meta-atoms.

Non-locality refers to the fact that the response of a meta-atom is not exclusively determined by its local properties but also by its interactions with distant (``non-local'') meta-atoms. Non-locality is the norm rather than an exception in electromagnetism, and more generally wave engineering, but it was often neglected or treated as a nuisance. More recently, however, non-local interactions between meta-atoms have been embraced to tailor the dispersion relations of diverse metamaterial platforms~\cite{shastri2022nonlocal,lemoult2013wave,chen2021roton,chennonlocal} and to achieve new functionalities for applications including analog signal processing~\cite{silva2014performing,kwon2018nonlocal,sol2022meta,sol2023reflectionless}, space compression~\cite{guo2020squeeze,reshef2021optic,chen2021dielectric} and the multi-functionality of metasurfaces~\cite{kamali2017angle,zhang2020controlling,overvig2020multifunctional}. 
Usually, non-locality is discussed in the canonical basis, e.g., in terms of direct interactions of far-apart meta-atoms. For instance, Ref.~\cite{chennonlocal} considered a 1D chain of meta-atoms and referred to direct interactions between spatially-nearest neighbors as ``local'' and direct interactions between beyond-spatially-nearest neighbors as ``non-local''. However, as stated above and as we will develop in the present paper, the role of the secondary meta-atoms (the ones not directly connected to any asymptotic scattering channel) and their direct interactions among each other and with the primary meta-atoms can be understood as direct non-local interactions between the primary meta-atoms. This representation of the metamaterial in a basis reduced to the primary meta-atoms is fully equivalent to the conventional representation in the canonical basis. 

Typically, a $\mathcal{P}$ symmetry of the metamaterial is apparent in the canonical basis and hence also in the reduced basis of primary meta-atoms. However, it is also possible that there is no $\mathcal{P}$ symmetry in the canonical basis but that there is one in the reduced basis of primary meta-atoms. Examples of ``hidden'' symmetries have been discovered in various electronic (or tight-binding) structures where they can give rise to ``non-accidental'' degeneracies~\cite{bayer2000hidden,hou2013hidden,bultinck2020ground,rontgen2021latent,tepliakov2023unveiling}. However, symmetries (often in combination with tuning) also play a pivotal role in wave scattering control, and being able to hide them from sight by covertly encoding them into complex coupling mechanisms between the meta-atoms may protect wave devices against reverse engineering and offer a route toward physical-layer secure communications. 

As an example of the role of symmetries in wave scattering control, let us consider the long-standing problem of reflectionless excitation of a scattering system. Wave devices with signal routing functionalities such as mode sorters and demultiplexers are deployed in larger nanophotonic or radio-frequency networks used to transfer information or energy. Being able to couple waves into such wave devices without any reflection not only prevents a loss of signal power, but more importantly avoids that reflected-power echoes within the network imperil non-linear components like lasers or power amplifiers. The reflectionless excitation of a generic arbitrarily complex scattering system through a subset of the connected asymptotic scattering channels is possible whenever the corresponding filtered version of its scattering matrix has a zero eigenvalue~\cite{sweeney2020theory}. In general, the continuous tunability of a system parameter is necessary to meet this condition~\cite{sweeney2020theory}, and if additional constraints on the specific frequency or routing functionalities exist, more tunable degrees of freedom are needed~\cite{imani2020perfect,sol2022meta,sol2023reflectionless}. If, however, the system has a parity-time ($\mathcal{PT}$) symmetry, then the system's scattering matrix is guaranteed to have zero eigenvalues at some frequencies without any tuning~\cite{gorbatsevich2016pt,gorbatsevich2017coalescence,sweeney2020theory,stone2020reflectionless,dhia2018trapped}. Moreover, a single symmetry-preserving continuous tuning parameter is sufficient to achieve the coalescence of two zero eigenvalues and their eigenvectors, giving rise to a reflectionless EP with broader lineshape. The easy access to reflectionless EPs~\cite{achilleos2017non,gorbatsevich2017unified,sweeney2019perfectly,sweeney2020theory,wang2021coherent,ferise2022exceptional}, which are distinct from resonant and scattering EPs, is important for broadening the bandwidth of (almost perfect) reflection suppression as well as for analog higher-order signal differentiation~\cite{sol2022meta,ferise2022exceptional}. Furthermore, the extreme sensitivity of EPs to detuning may also be the basis of new sensor concepts. Given a metamaterial with hidden $\mathcal{P}$ symmetry where absorption (and hence the ensuing $\mathcal{T}$ symmetry breaking) is sufficiently small, one can \textit{covertly} implement the described $\mathcal{PT}$-symmetry-based scattering control.

In this paper, building on recent theoretical work~\cite{rontgen2022hidden,rontgen2023scattering}, we develop a scattering theory of hidden $\mathcal{P}$ symmetries in generic arbitrarily complex wave systems. Then, for our experiments, we focus on cable-network metamaterials~\cite{chennonlocal} which are transmission-line networks also known as ``quantum graphs''~\cite{kottos1997quantum,kottos2003quantum,hul2004experimental}. Therein, the junctions (vertices) are non-resonant meta-atoms and the cables (bonds) are the direct interactions between the meta-atoms. Various kinds of waveguide networks were recently also explored in contexts such as nanophotonic lasing~\cite{gaio2019nanophotonic}, non-Abelian topological charges, edge states and braiding~\cite{guo2021experimental,chen2022classical} or topology-protected wave devices~\cite{zhang2021superior}. The reconfigurable ``plug-and-play'' nature of cable-network metamaterials makes them an ideal platform to prototype covert scattering control, and, moreover, they are of direct relevance to wireline communications networks. Besides the hidden $\mathcal{P}$ symmetry, our covert scattering control requires tuning mechanisms, one that breaks and one that preserves the hidden $\mathcal{P}$ symmetry. 
Based on these tools, first, we demonstrate a scheme for physical-layer secure wireline communications in which the hidden $\mathcal{P}$ symmetry constitutes an advantage for the legitimate receiver over the eavesdropper. 
Then, second, within the approximation of negligible absorption, we access the regime of $\mathcal{PT}$ symmetry and demonstrate the tuning to reflectionless EPs in metamaterials \textit{without} mirror symmetry. We also observe quasi-bound states in the continuum (qBICs)~\cite{hsu2016bound}.
Finally, we extend this covert scattering control to metamaterials with non-reciprocal interactions between meta-atoms, and we observe reflectionless EPs in a non-recirocal system for the first time.

\clearpage

\section{Generalities}
\label{sec0}

To start, we establish the theoretical foundation of the intuitive understanding of hidden $\mathcal{P}$ symmetry in non-local metamaterials outlined in the introduction. Earlier work studied hidden $\mathcal{P}$ symmetry in quantum graphs~\cite{rontgen2022hidden,rontgen2023scattering}, and has some overlap with this section, although we present a more general formulation that clarifies the importance of the coupling between the primary meta-atoms and the asymptotic scattering channels; our formulation also reveals the generality of the theory, with obvious extensions to other wave systems described by representations such as the discrete-dipole approximation~\cite{de1998point}, temporal coupled-mode theory~\cite{suh2004temporal} and, most generally, Weidenmüller's generalization of Breit-Wigner theory~\cite{beenakker1997random}.

\subsection{Wave-Operator Representation of the Scattering Matrix}

Througout this paper, we work in the time-harmonic regime. For any finite arbitrary linear scattering system connected to $m$ (mutually orthogonal) asymptotic scattering channels, the scattering matrix $\mathbf{S}\in \mathbb{C}^{m \times m}$ can be approximately formulated in a microscopic sense as~\cite{fano1961effects,viviescas2003field,rotter2017light,sweeney2020theory}

\begin{equation}
    \mathbf{S} = \mathbf{I} - 2 i \mathbf{W}^\dagger \frac{1}{\mathbf{H} +i \mathbf{W} \mathbf{W}^\dagger  } \mathbf{W},
    \label{eq_s_matrix_gen}
\end{equation}

\noindent where $\mathbf{H} = \mathbf{A_0} - \mathbf{\Delta}$. $\mathbf{A_0}\in \mathbb{C}^{n \times n}$ is the wave operator in the closed system, where we assume a sufficiently high-resolution discretization of Maxwell’s equations into $n$ voxels. $\mathbf{W}\in \mathbb{C}^{n \times m}$ is the matrix describing the coupling between each voxel and each asymptotic scattering channel. $\mathbf{\Delta}$ is the principal value of $\int{\mathrm{d}\omega' \frac{ \mathbf{W}  \mathbf{W}^\dagger  }{\omega' - \omega}}$. All involved matrices except $\mathbf{I}$ depend in general on the angular frequency $\omega$ but we do not explicitly print this dependence for conciseness. 
The terms $\mathbf{\Delta}$ and $-i\mathbf{W} \mathbf{W}^\dagger$ are the Hermitian and anti-Hermitian parts of the self-energy that originates from the boundary between the finite closed system and the exterior asymptotic region, and they describe, respectively, frequency shifts and damping terms that arise due to coupling the closed system to the outside world's continuum. The formulation in Eq.~(\ref{eq_s_matrix_gen}) originates from nuclear scattering theory~\cite{mahaux1969shell,beenakker1997random} and does not constitute an exact computational method for scattering calculations~\cite{zhang2020quasinormal}.

While the theory presented in Sec.~\ref{sec_geometricSymm} and Sec.~\ref{sec_latentSymm} applies to any generic wave system described by Eq.~(\ref{eq_s_matrix_gen}), the subsequent experiments presented in the current paper are based specifically on cable-network metamaterials. Although, of course, the general formulation in the microscopic sense from Eq.~(\ref{eq_s_matrix_gen}) applies, it is more convenient to consider an analogous macroscopic formulation. Therein, each junction is a (non-resonant) meta-atom that plays the role of a voxel in the microscopic formulation, and it is similarly directly coupled to each of its closest neighbors, in the macroscopic case via cables. Our cable-network metamaterials are also known as Neumann quantum graphs, and Refs.~\cite{texier2001scattering,kottos2003quantum,kuchment2003quantum} derived a macroscopic formulation that takes exactly the form of Eq.~(\ref{eq_s_matrix_gen}). In this specific case, the formulation from Eq.~(\ref{eq_s_matrix_gen}) constitutes an exact expression without any truncations that can be used for computational scattering calculations~\cite{kottos2003quantum}. 
Therein, under the assumption that each asymptotic scattering channel is non-dispersively coupled to exactly one meta-atom,
\begin{equation}
    H_{i,j} = 
\begin{cases}
    -\sum_{l \neq i} C_{i,l} \ \mathrm{cot}(k L_{i,l}) & \text{if } i = j.\\
    C_{i,j} \ \mathrm{exp}(-iA_{i,j}L_{i,j}) \ \mathrm{csc}(k L_{i,j})              & \text{otherwise.}
\end{cases}
\label{eq_quantumgraphsmatrix}
\end{equation}

\noindent Here, $C_{i,j}$ is unity if the $i$th and $j$th meta-atoms are directly connected, and zero otherwise. $L_{i,j}$ is the length of the cable connecting the $i$th and $j$th meta-atoms, and $A_{i,j}$ is a magnetic vector potential acting on that bond (if there is no magnetic vector potential, $A_{i,j}=0$). $k$ is the wavenumber. 
The diagonal and off-diagonal entries of $\mathbf{H}$ are also referred to as self-interaction or on-site potential and hoping, respectively. The $(i,j)$th entry of $\mathbf{W}$ is unity if the $i$th meta-atom is connected to the $j$th asymptotic scattering channel, and zero otherwise. 

So far, we describe the wave operator in the canonical representation. This representation has an obvious interpretation as a graph. For the case of our cable-network metamaterial, this interpretation is already cemented into the frequently used alternative term ``quantum graph'', but this graph interpretation applies to wave systems in general. For instance, recently, it has been discussed more explicitly in the case of coupled-dipole systems~\cite{scali2023graph,rabault2023nonlinearity}. 
In the following Sec.~\ref{sec_geometricSymm}, we discuss $\mathcal{P}$ symmetry in this canonical representation. Meanwhile, it is important to note that other equivalent representations exist. For instance, one could change to the modal basis (we refer the interested reader to Ref.~\cite{zhang2020quasinormal} where a rigorous ab initio quasinormal coupled mode theory is derived). A representation of importance for the present paper is one that is reduced to the primary meta-atoms. In Sec.~\ref{sec_latentSymm}, we derive such a representation in order to discuss how $\mathcal{P}$ symmetry can be covertly encoded into the topology of the secondary meta-atoms.

\subsection{$\mathcal{P}$-Symmetry in the Canonical Representation}
\label{sec_geometricSymm}

$\mathcal{P}$ symmetry of the wave operator in the canonical basis, e.g., a mirror-symmetry, is \textit{not} a sufficient condition to guarantee $\mathcal{P}$ symmetry of the scattering matrix. It is further required that the way in which the wave operator is coupled to the asymptotic scattering channels preserves the wave operator's $\mathcal{P}$ symmetry. This implies the need for a bisected partition of the channels, the most natural one being a division into ``left'' and ``right''~\cite{sweeney2020theory}. More rigorously, we can formulate these two conditions as follows: if $\mathcal{\hat{P}}_n\mathbf{A_0}=\mathbf{A_0}\mathcal{\hat{P}}_n$ \textit{and} $\mathbf{W}\mathcal{\hat{P}}_m = \mathcal{\hat{P}}_n\mathbf{W}$, then $\mathcal{\hat{P}}_m\mathbf{S}=\mathbf{S}\mathcal{\hat{P}}_m$ (see Supplementary Note I). Note that these conditions are compatible with $\mathcal{P}$-symmetry-preserving absorption mechanisms such as homogeneous absorption. 

Given that $\mathbf{A_0}$ commutes with $\mathcal{\hat{P}}_n$, the eigenstates of the closed system must also be eigenstates of $\mathcal{\hat{P}}_n$ and as such have definite parity. In the case of a spatial mirror symmetry, this means that the value of an eigenstate $\phi$ at a position $\mathbf{r}$ is $\pm1$ times the value of the same eigenstate at the mirror-symmetric position $\mathcal{R}(\mathbf{r})$: $\phi(\mathcal{R}(\mathbf{r})) = \pm \phi(\mathbf{r})$. For a mirror-symmetric system with $m=2$ (one ``left'' and one ``right'' channel), it follows that $S_{11}(\omega_0) = S_{22}(\omega_0)$ (which we refer to as ``equi-reflection'' property) and $S_{21}(\omega_0) = S_{12}(\omega_0)$ (which is known as reciprocity).

\subsection{$\mathcal{P}$-Symmetry in the Reduced Basis of Primary Meta-Atoms}
\label{sec_latentSymm}

Now, let us consider the scenario in which a subset $\mathcal{\bar{S}}$ of $v<n$ voxels or meta-atoms are not directly coupled to any asymptotic scattering channel. In our terminology, the metamaterial then contains $n-v$ primary and $v$ secondary meta-atoms. 
We can then write $\mathbf{A_0}$ and $\mathbf{W}$ in block form:

\begin{subequations}
\begin{equation}
\mathbf{A_0} = \begin{bmatrix} 
	\mathbf{A_{0,\mathcal{S}\mathcal{S}}}  & \mathbf{A_{0,\mathcal{S}\mathcal{\bar{S}}}} \\
	\mathbf{A_{0,\mathcal{\bar{S}}\mathcal{S}}}  & \mathbf{A_{0,\mathcal{\bar{S}}\mathcal{\bar{S}}}} \\
\end{bmatrix}
\end{equation}
\label{A_block}
\begin{equation}
\mathbf{W} = \begin{bmatrix} 
	\mathbf{W_{\mathcal{S}}}   \\
	\mathbf{0}   \\
\end{bmatrix}
\end{equation}
\label{W_block}
\end{subequations}

\noindent where $\mathbf{A_{0,\mathcal{S}\mathcal{S}}}\in \mathbb{C}^{(n-v) \times (n-v)}$, $\mathbf{A_{0,\mathcal{S}\mathcal{\bar{S}}}}\in \mathbb{C}^{(n-v) \times v}$, $\mathbf{A_{0,\mathcal{\bar{S}}\mathcal{S}}}\in \mathbb{C}^{v \times (n-v)}$, $\mathbf{A_{0,\mathcal{\bar{S}}\mathcal{\bar{S}}}}\in \mathbb{C}^{v \times v}$, $\mathbf{W_{\mathcal{S}}}\in \mathbb{C}^{(n-v) \times m}$, and $\mathbf{0}$ is of dimensions 
$v \times m$. Then,

\begin{equation}
\mathbf{H} + i\mathbf{W}\mathbf{W}^\dagger= \mathbf{G} = \begin{bmatrix} 
	\mathbf{G_{\mathcal{S}\mathcal{S}}}  & \mathbf{G_{\mathcal{S}\mathcal{\bar{S}}}} \\
	\mathbf{G_{\mathcal{\bar{S}}\mathcal{S}}}  & \mathbf{G_{\mathcal{\bar{S}}\mathcal{\bar{S}}}}\end{bmatrix}  = \begin{bmatrix} 
	\mathbf{A_{0,\mathcal{S}\mathcal{S}}}  & \mathbf{A_{0,\mathcal{S}\mathcal{\bar{S}}}} \\
	\mathbf{A_{0,\mathcal{\bar{S}}\mathcal{S}}}  & \mathbf{A_{0,\mathcal{\bar{S}}\mathcal{\bar{S}}}} \\
\end{bmatrix} - \left( \begin{bmatrix}\mathbf{\Delta_{\mathcal{S}\mathcal{S}}}  & \mathbf{0} \\
	\mathbf{0}  & \mathbf{0} 
\end{bmatrix} -i  \begin{bmatrix}\mathbf{W_{\mathcal{S}}}\mathbf{W_{\mathcal{S}}}^\dagger  & \mathbf{0} \\
	\mathbf{0}  & \mathbf{0} 
\end{bmatrix}\right),    
\end{equation}

\noindent where $\mathbf{\Delta_{\mathcal{S}\mathcal{S}}}$ is the principal value of $\int{\mathrm{d}\omega' \frac{ \mathbf{W_{\mathcal{S}}}  \mathbf{W_{\mathcal{S}}}^\dagger  }{\omega' - \omega}}$. Now, to obtain $\mathbf{S}$ we must invert the $2\times 2$ block matrix $ \mathbf{G}$. Given Eq.~(\ref{W_block}b), only the top left block of $ \mathbf{G}^{-1}$ matters and we obtain

\begin{equation}
    \mathbf{S} = \mathbf{I} - 2 i \mathbf{W_{\mathcal{S}}}^\dagger \frac{1}{\mathbf{G_{\mathcal{S}\mathcal{S}}}-\mathbf{G_{\mathcal{S}\mathcal{\bar{S}}}}\mathbf{G_{\mathcal{\bar{S}}\mathcal{\bar{S}}}}^{-1}\mathbf{G_{\mathcal{\bar{S}}\mathcal{{S}}}}}     \mathbf{W_{\mathcal{S}}}
    = \mathbf{I} - 2 i \mathbf{W_{\mathcal{S}}}^\dagger \frac{1}{\mathbf{\tilde{A}_0} - (\mathbf{\Delta_{\mathcal{S}\mathcal{S}}} -i  \mathbf{W_{\mathcal{S}}} \mathbf{W_{\mathcal{S}}}^\dagger ) } \mathbf{W_{\mathcal{S}}},
\label{eq_s_matrix_effective_wave_operator}
\end{equation}

\noindent where we have defined an effective wave operator 

\begin{equation}
    \mathbf{\tilde{A}_0} = \mathbf{A_{0,\mathcal{S}\mathcal{S}}}-\mathbf{A_{0,\mathcal{S}\mathcal{\bar{S}}}}\mathbf{A_{0,\mathcal{\bar{S}}\mathcal{\bar{S}}}}^{-1}\mathbf{A_{0,\mathcal{\bar{S}}\mathcal{{S}}}}.
    \label{effective_wave_operator}
\end{equation}

\begin{figure*}[htbp]
    \centering
    \includegraphics[width=\columnwidth]{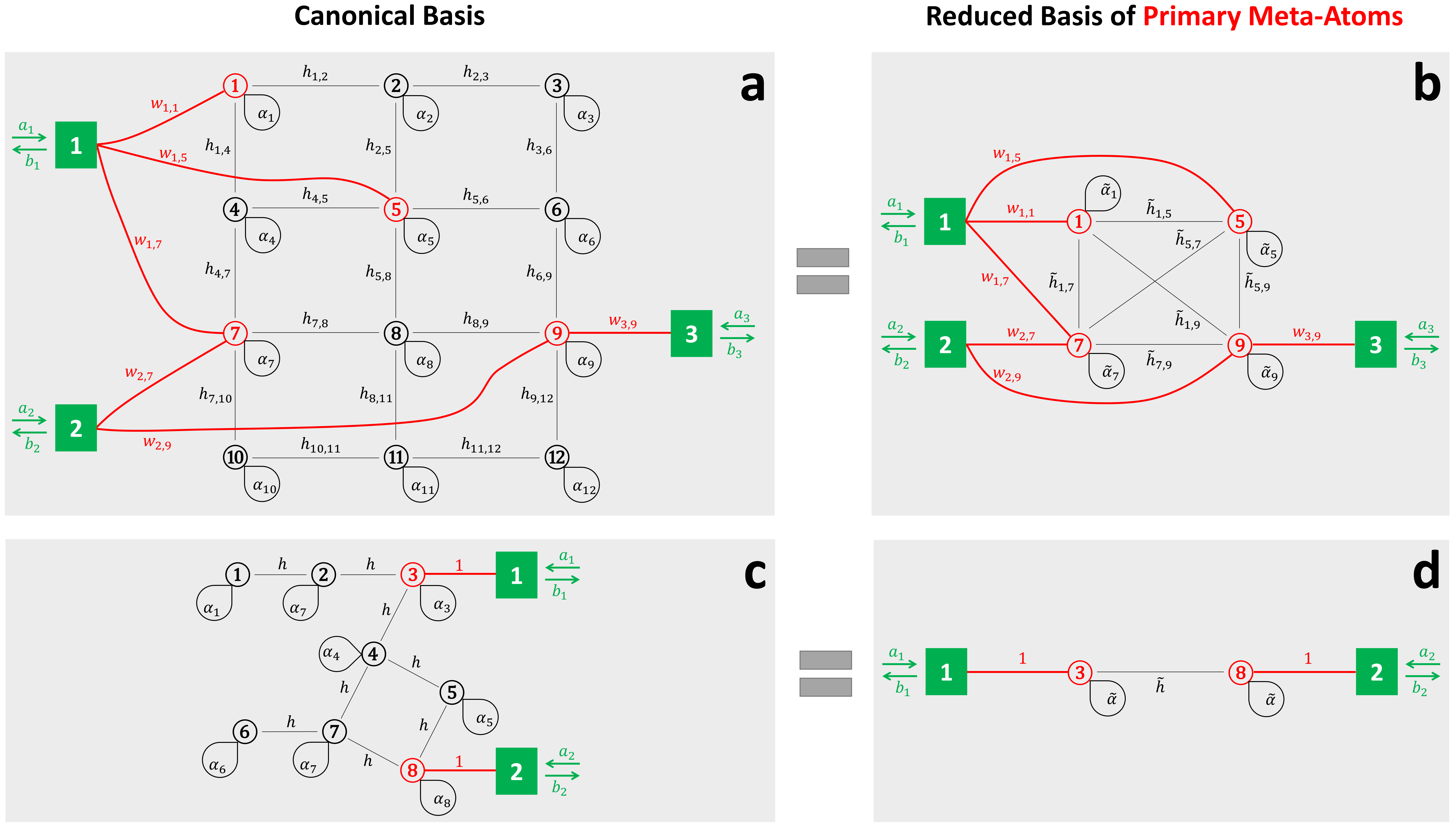}
    \caption{\textbf{General Principle of Reduced-Basis Representations and Hidden $\mathcal{P}$ Symmetry.} (a) Conceptual sketch of a generic example scattering problem involving three asymptotic scattering channels and 12 meta-atoms. The $i$th meta-atom has a self-interaction $\alpha_i$ and a reciprocal interaction $h_{i,j}$ with its direct neighbors. The $i$th asymptotic scattering channel is coupled to the $j$th meta-atom with a complex-valued weight $w_{i,j}$ (only non-zero weights are shown). Primary meta-atoms have direct contact with one or more asymptotic scattering channel(s) and are highlighted in red. Lines in this figure symbolically represent direct coupling. In the case of a cable-network metamaterial, direct coupling is implemented via cables that look ``line-like'' and the asymptotic scattering channels are monomodal waveguides. In general, however, the coupling mechanisms can be more complex. For instance, if a shaped wavefront illuminates a nanophotonic structure, the free-space asymptotic scattering channel couples with different complex-valued weights to multiple internal scattering entities depending on how the wavefront pattern overlaps with them.
    (b) Equivalent representation of the metamaterial from (a) in the reduced basis of primary meta-atoms. Here, the $i$th primary meta-atom has a self-interaction $\tilde\alpha_i$ and has an interaction $\tilde h_{i,j}$ with the $j$th primary meta-atom. (c,d) Example of a metamaterial with a hidden $\mathcal{P}$ symmetry that is absent in the canonical basis (c) but evident in the reduced basis of primary meta-atoms (d).}
    \label{fig1}
\end{figure*}

\noindent Physically, as illustrated in Fig.~\ref{fig1}, an equivalent description of our system consists hence in interpreting only the subset $\mathcal{S}$ of voxels as our system's internal scattering entities, and the remaining subset $\mathcal{\bar{S}}$ of voxels as complicated additional coupling mechanisms between the scattering entities included in $\mathcal{S}$. In other words, for our macroscopic description of the cable-network metamaterial, an equivalent description of the system is possible in the reduced representation of the primary meta-atoms wherein the secondary meta-atoms are treated as additional coupling mechanisms between the primary meta-atoms. 
Similar calculations of effective wave operators like Eq.~(\ref{effective_wave_operator}) were also presented within the more limited scopes of tight-binding network engineering~\cite{longhi2016non} and isospectral graph reduction~\cite{bunimovich2014isospectral}, however, without the complete scattering calculation from Eq.~(\ref{eq_s_matrix_effective_wave_operator}).

Analogous to Sec.~\ref{sec_geometricSymm}, we conclude that if $\mathcal{\hat{P}}_{n-v}\mathbf{\tilde{A}_0}=\mathbf{\tilde{A}_0}\mathcal{\hat{P}}_{n-v}$ \textit{and}
     $\mathbf{W_{\mathcal{S}}}\mathcal{\hat{P}}_m = \mathcal{\hat{P}}_{n-v}\mathbf{W_{\mathcal{S}}}$, then $\mathcal{\hat{P}}_m\mathbf{S}=\mathbf{S}\mathcal{\hat{P}}_m$. 
These conditions are compatible with absorption mechanisms that preserve the effective wave operator's parity, for instance, homogeneous absorption. 
Unlike Sec.~\ref{sec_geometricSymm}, in which the parity of the wave operator was domain-wise (excluding the trivial case of a mirror-symmetric point), here the parity is restricted to the subset $\mathcal{S}$ which may be point-wise, for instance, in the most extreme case of $n-v=m=2$.  Given a system with hidden $\mathcal{P}$ symmetry, it is always possible to extend the domain defined by $\mathcal{S}$ by symmetrically coupling additional clusters of meta-atoms to the primary meta-atoms. 
It also follows that the eigenstates of $\mathbf{\tilde{A}_0}$ must have definite parity. In other words, the eigenstates of $\mathbf{A_0}$ have definite parity, but \textit{only} within the subset $\mathcal{S}$ of the system's internal scattering entities. In Supplementary Note III, we report experimental measurements to confirm this definite parity of the eigenmodes.

The conditions $\mathcal{\hat{P}}_{n-v}\mathbf{\tilde{A}_0}=\mathbf{\tilde{A}_0}\mathcal{\hat{P}}_{n-v}$ \textit{and} $\mathbf{W_{\mathcal{S}}}\mathcal{\hat{P}}_m = \mathcal{\hat{P}}_{n-v}\mathbf{W_{\mathcal{S}}}$ can be satisfied by scattering systems that do \textit{not} satisfy $\mathcal{\hat{P}}_n\mathbf{A_0}=\mathbf{A_0}\mathcal{\hat{P}}_n$ \textit{and} $\mathbf{W}\mathcal{\hat{P}}_m = \mathcal{\hat{P}}_n\mathbf{W}$. 
For instance, Refs.~\cite{rontgen2022hidden,rontgen2023scattering} theoretically identified concrete examples thereof for the case of $m=n-v=2$ and all bonds being reciprocal. In these examples, of which one is provided in Fig.~\ref{fig1}(c,d), all meta-atoms are non-resonant and all pairs of directly connected meta-atoms have the same interaction $h$. Then, the reduced representation turns out to satisfy the first condition $\mathcal{\hat{P}}_{n-v}\mathbf{\tilde{A}_0}=\mathbf{\tilde{A}_0}\mathcal{\hat{P}}_{n-v}$. Moreover, the second condition $\mathbf{W}\mathcal{\hat{P}}_m = \mathcal{\hat{P}}_n\mathbf{W}$ is trivially satisfied because any given primary meta-atom is (non-dispersively) directly coupled to exactly one asymptotic scattering channel.

\section{Application to Physical-Layer Secure Communications}

As a first example of the technological relevance of the intriguing concept of hidden $\mathcal{P}$ symmetry, we demonstrate in this section that covert scattering control, built off a hidden $\mathcal{P}$ symmetry, is ideally suited to establish physical-layer secure wireline communications.

\begin{figure*}
    \centering
    \includegraphics[width=\columnwidth]{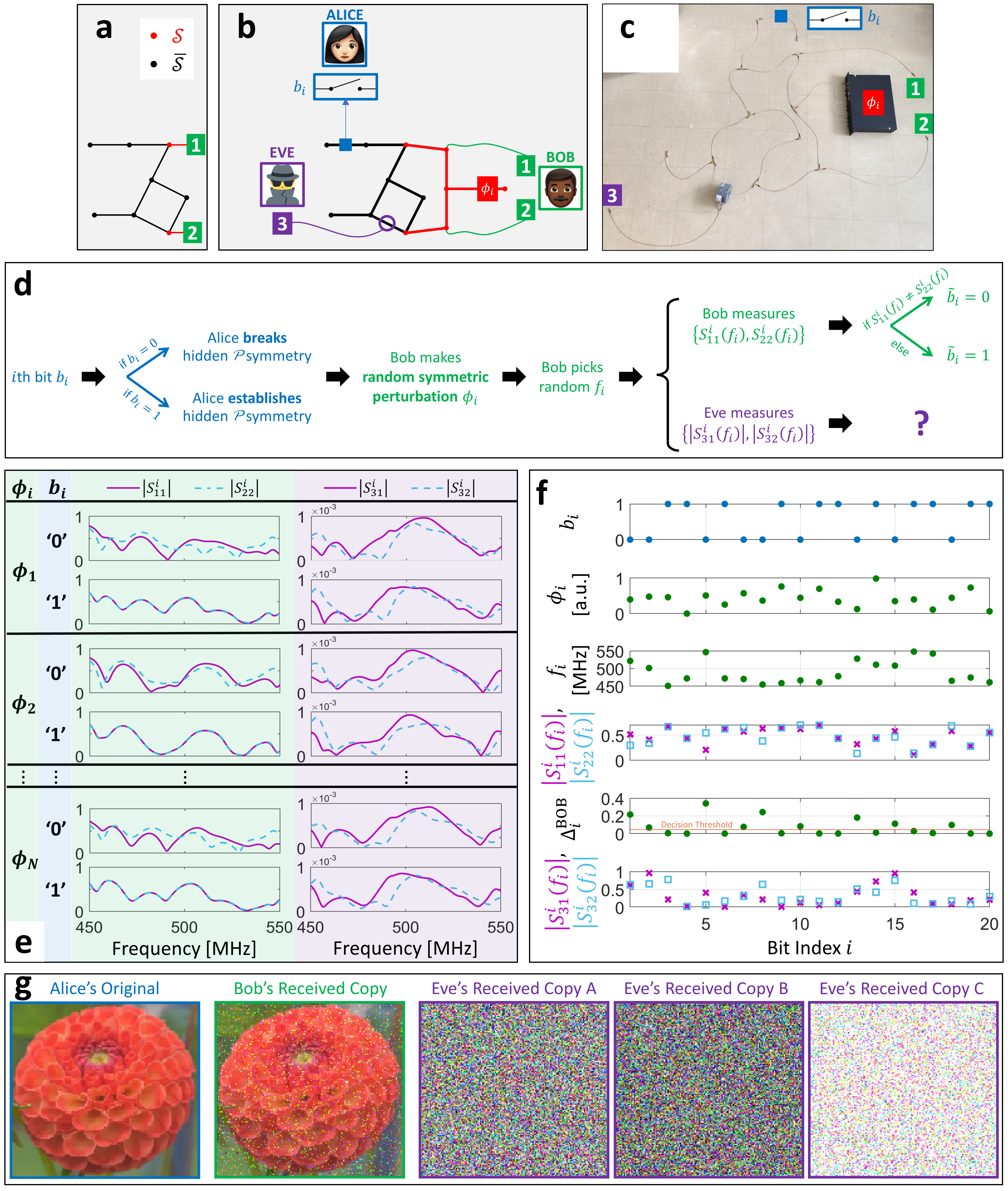}
    \caption{\textbf{Physical-Layer Secure Communication Built Off Hidden $\mathcal{P}$ Symmetry.} 
    (a) Wireline network with hidden $\mathcal{P}$ symmetry of its primary (red) meta-atoms. 
    (b) Addition of hidden-$\mathcal{P}$-symmetry-preserving tuning mechanism (Bob's phase shifter $\phi_i$) and hidden-$\mathcal{P}$-symmetry-breaking tuning mechanism (Alice's switch $b_i$) to the network from (a). In addition, Eve non-invasively wire-
    }
    \label{fig2}
\end{figure*}

\setcounter{figure}{1}    
\begin{figure*}[h]
    \centering
    \includegraphics[width=0.0001\columnwidth]{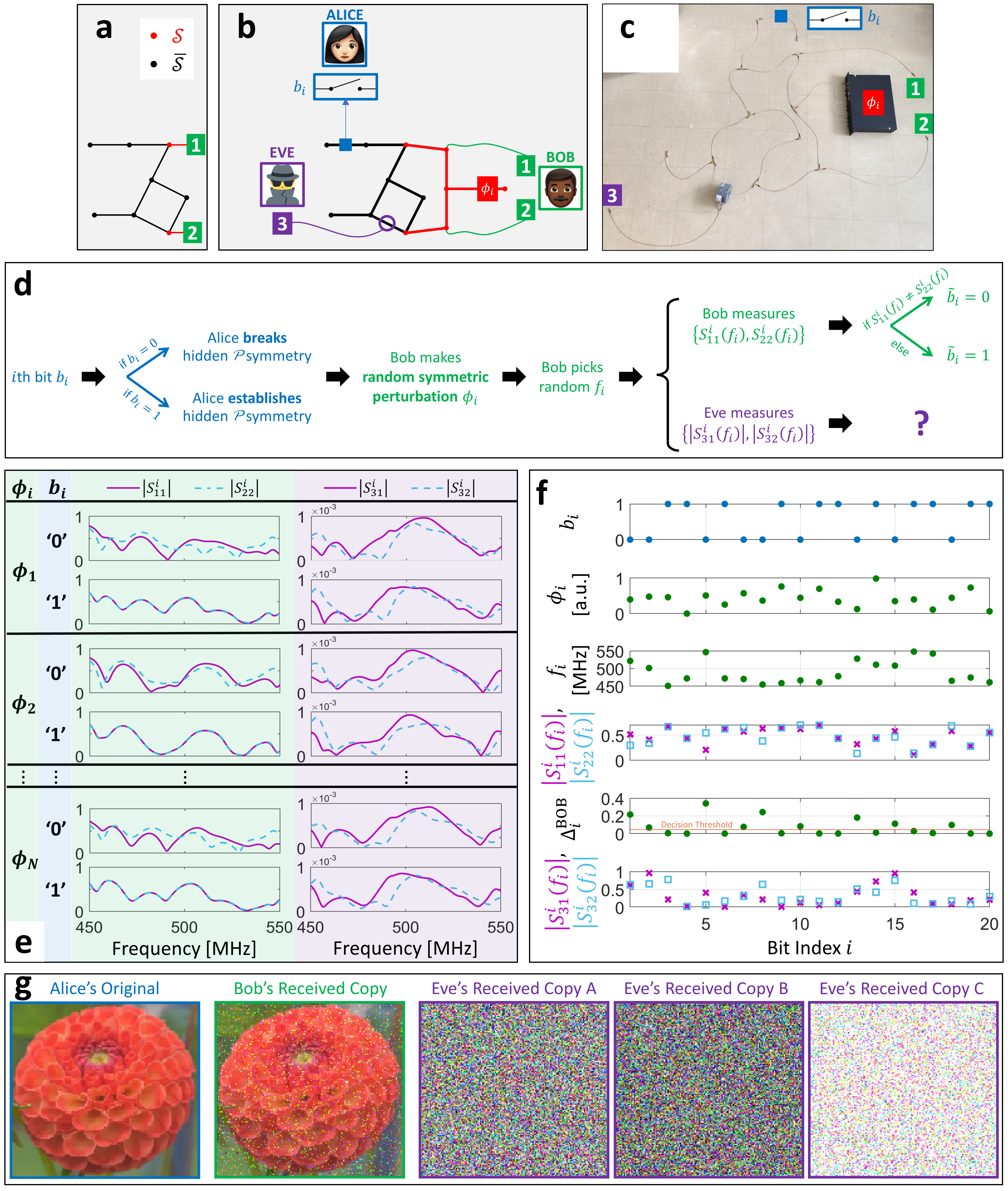}
    \caption{[CONTINUED] taps the network. 
    (c) Photographic image of the setup described in (b). 
    (d) Physical-layer secure communications protocol. 
    (e) Measured reflection (Bob) and transmission (Eve) magnitude spectra for different settings of the tuning mechanisms $\phi_i$ and $b_i$. 
    (f) Example of Alice's bit stream $b_i$, Bob's choices of $\phi_i$ and $f_i$, Bob's measurements of $S_{11}^i(f_i)$ and $S_{22}^i(f_i)$ (only magnitude is shown), the difference $\Delta_i^{\mathrm{BOB}}$ between Bob's measurements to decide if there is a hidden $\mathcal{P}$ symmetry, and Eve's measurements $|S_{31}^i(f_i)|$ and $|S_{32}^i(f_i)|$.
    (g) Secure transfer of a photographic image of a \textit{Dahlia `Bantling'} flower taken in Rennes, France. Eve cannot decode the image using Bob's decoder (A), a K-means algorithm (B), or an autoencoder algorithm (C).}
    \label{fig2cont}
\end{figure*}

For concreteness, let us consider the metamaterial with hidden $\mathcal{P}$ symmetry from Fig.~\ref{fig1}(c) that is connected to $m=2$ asymptotic scattering channels. In a simple communication protocol leveraging the hidden $\mathcal{P}$ symmetry, the sender (Alice) could encode the confidential data stream in an on-off-keying (OOK) manner by using a hidden-$\mathcal{P}$-symmetry-breaking tuning mechanism (e.g., a switch that alters the topology of the secondary meta-atoms), and the legitimate receiver (Bob) would decode the information by checking if he observes the same reflection coefficient on ports 1 and 2. This is a so-called backscatter-communication scheme because Alice modulates a wave generated by Bob \footnote{Backscatter-communication schemes are increasingly popular in \textit{wireless} communications~\cite{weinstein2005rfid,liu2013ambient,zhao2020metasurface} but in the present work we are concerned with a backscatter-communication scheme for \textit{wireline} networks.}. 
We will further assume that the eavesdropper Eve cannot access the subdomain $\mathcal{S}$ of the metamaterial, such that she can never check the parity therein. Moreover, Eve cannot make any invasive measurements by strongly coupling additional asymptotic scattering channels to the subdomain $\mathcal{\bar{S}}$ because the resulting perturbation of the metamaterial would destroy the hidden $\mathcal{P}$ symmetry in the subdomain $\mathcal{S}$, such that Bob would immediately detect the presence of Eve. Hence, Eve can only non-invasively monitor the transmission magnitudes from ports 1 and 2 to her port 3 that she can couple very weakly (non-invasively) to some part of the subdomain $\mathcal{\bar{S}}$. Whereas Bob can measure the complex-valued reflection coefficients $S_{11}$ and $S_{22}$, Eve can only measure the magnitude of the transmissions  $|S_{31}|$ and $|S_{32}|$ because she lacks synchronization with Bob's source. Given the weak coupling of Eve's port 3 to the metamaterial, besides being the only one who can check whether the metamaterial has a hidden $\mathcal{P}$ symmetry, Bob will have a strong signal-to-noise ratio (SNR) advantage over Eve. However, we will focus on the former in the following by assuming very low noise levels for both Bob and Eve.

Nonetheless, the scheme outlined so far is not yet secure. Eve is not given a chance to check if there is a hidden $\mathcal{P}$ symmetry in the subdomain $\mathcal{S}$, but she can notice that her measurements only alternate between two possible pairs of transmission spectra: $\{|S_{31}^\mathcal{P}(f)|,|S_{32}^\mathcal{P}(f)|\}$ and $\{|S_{31}^{\bar{\mathcal{P}}}(f)|,|S_{32}^{\bar{\mathcal{P}}}(f)|\}$. With sufficient dynamic range, Eve could easily distinguish these two cases and decode the OOK of Alice's confidential message. To close this loophole, Bob can deploy a hidden-$\mathcal{P}$-symmetry-preserving tuning mechanism. While the setting of the latter will not affect the presence or absence of the hidden $\mathcal{P}$ symmetry in the subdomain $\mathcal{S}$, it will confuse Eve. Specifically, Eve will be confronted with a new pair of transmission spectra for every symbol and can no longer decode the OOK because she cannot distinguish whether a change in her pair of measured transmission spectra originates from Alice's hidden-$\mathcal{P}$-breaking tuning that encodes the confidential information or Bob's hidden-$\mathcal{P}$-preserving tuning that merely serves to confuse Eve. To implement a hidden-$\mathcal{P}$-preserving tuning mechanism, Bob can simply connect a programmable phase shifter symmetrically to ports 1 and 2, as shown in Fig.~\ref{fig2}(b). 

The final secure communication protocol is summarized in Fig.~\ref{fig2}(d). Example measurements of Bob's reflection spectra and Eve's transmission spectra for different settings of Alice's hidden-$\mathcal{P}$-breaking tuning mechanism (denoted by $b_i$ because it encodes the confidential data steam) and Bob's hidden-$\mathcal{P}$-preserving tuning mechanism (denoted by $\phi_i$) are shown in Fig.~\ref{fig2}(e). The presence or absence of hidden $\mathcal{P}$ symmetry is directly obvious on Bob's data, irrespective of $\phi_i$. In contrast, Eve's data slightly varies with both $b_i$ and $\phi_i$. We further illustrate this in  Fig.~\ref{fig2}(f) for an example bit stream. For each symbol, Bob randomly picks a value for $\phi_i$ as well as the frequency $f_i$ at which he probes his two reflection coefficients. By comparing $\Delta_i^{\mathrm{BOB}} = | S_{11}^i(f_i) - S_{22}^i(f_i) |$ to a decision threshold, Bob easily decodes Alice's message. In contrast, Eve's measured data has no apparent relation to Alice's message. Even if Eve knows the principle of this physical-layer secure communication scheme and tries to transpose Bob's decoding scheme to her own data, she cannot decode Alice's message because the transmission magnitudes from ports 1 and 2 to port 3 are in general \textit{not} equal even if there is a hidden $\mathcal{P}$ symmetry in $\mathcal{S}$. Eve could also attempt to use other decoders like a K-means or autoencoder clustering algorithm to somehow group her received data into ``0'' and ``1'' symbols (see Supplemenary Note IV for details). However, as we illustrate for the transfer of an image in Fig.~\ref{fig2}(d), none of these techniques is successful. Meanwhile, Bob receives an almost flawless copy of the Alice's original image. The few imperfections arise when Bob chooses a frequency $f_i$ for which the difference between two reflection coefficients is accidentally small (below the decision threshold) in the case of no hidden $\mathcal{P}$ symmetry. Increasing the number of utilized frequencies per symbol could overcome this issue.

The complexity and size of the metamaterial with hidden $\mathcal{P}$ symmetry, as well as the number of legitimate receivers, can be scaled up by suitably combining metamaterials with hidden $\mathcal{P}$ symmetry~\cite{morfonios2021cospectrality}. 
Ultimately, the physical-layer security originates from an advantage (built upon the hidden $\mathcal{P}$ symmetry) for the legitimate receiver over the eavesdropper~\cite{shannon1949communication,wyner1975wire,leung1978gaussian,poor2017wireless}. Unlike Ref.~\cite{imani2020perfect}, in which Alice required a multitude of programmable meta-atoms to achieve wireless physical-layer secure communication based on imposing (or not) a perfect-absorption condition on Bob's port, in the present work Alice only requires a single hidden-$\mathcal{P}$-symmetry breaking tuning knob thanks to the broadband nature of the hidden $\mathcal{P}$ symmetry.

\clearpage

\section{Reflectionless Exceptional Points without Mirror Symmetry}

Let us now consider the approximation of negligible absorption in the tunable scattering system with hidden $\mathcal{P}$ symmetry from the previous section. Then, the system also has a hidden $\mathcal{PT}$ symmetry, enabling the \textit{covert} implementation of the $\mathcal{PT}$-symmetry-based wave control discussed in the introduction. 
In this section, we demonstrate such use of a hidden $\mathcal{PT}$ symmetry in combination with a hidden-$\mathcal{PT}$-symmetry-preserving tuning mechanism to implement \textit{reverse-engineering-resilient} access to reflectionless EPs, as well as quasi bound states in the continuum. As we explain below, the measurable absorption in our experimental system breaks exact $\mathcal{PT}$ symmetry such that we can only observe these features approximately, but future implementations of our work with standard complementary metal–oxide–semiconductor (CMOS) technology can readily compensate the attenuation with suitable gain mechanisms~\cite{cao2022fully}. Alternatively, complex-frequency excitations can provide ``virtual'' gain in a passive system like ours~\cite{li2020virtual}. 

\begin{figure*}
    \centering
    \includegraphics[width=0.8\columnwidth]{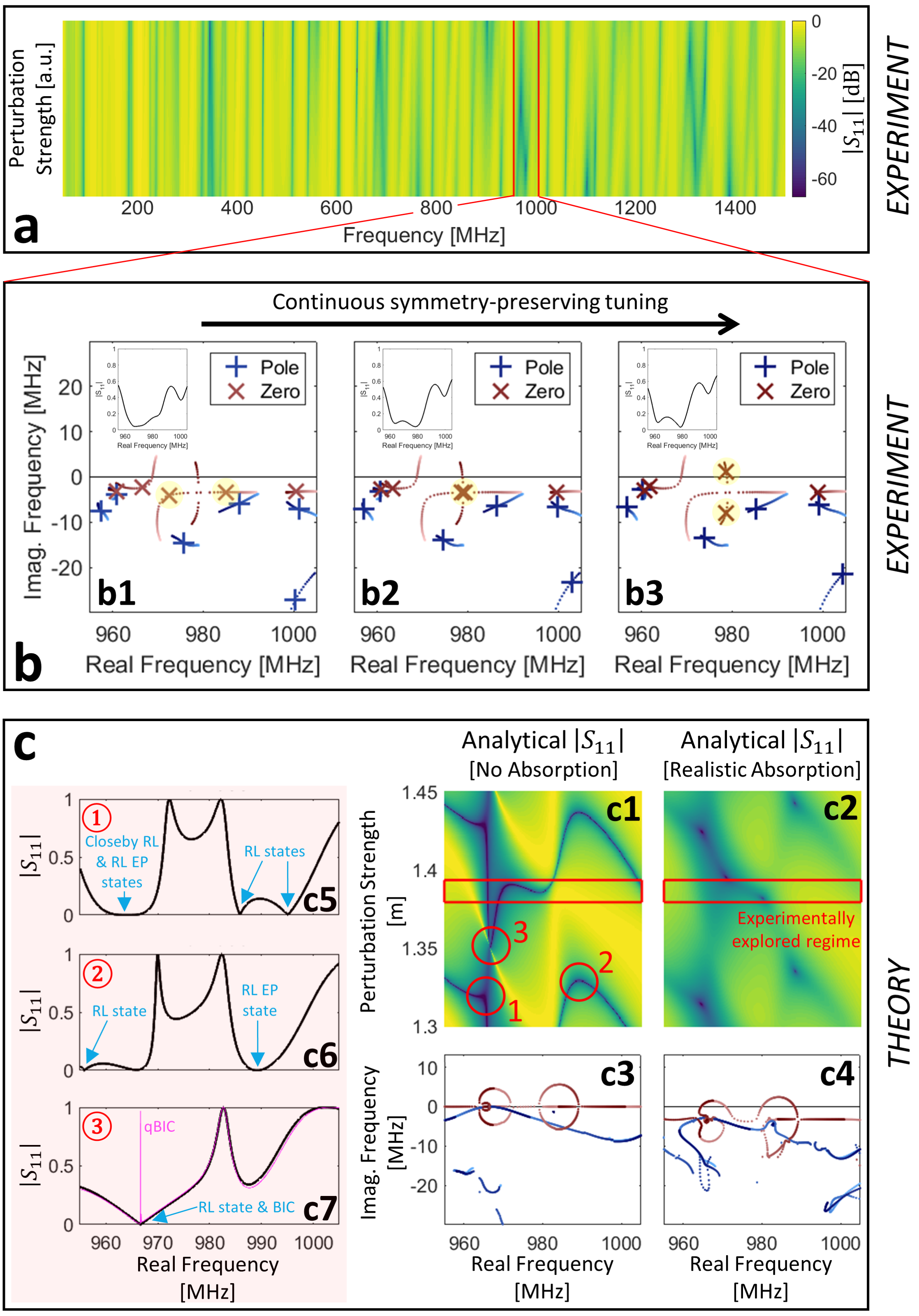}
    \caption{\textbf{Reflectionless Exceptional Points without Mirror Symmetry.} 
    (a) Experimentally measured perturbation-frequency map of $|S_{11}|$. The setup is that from Fig.~\ref{fig2}(c) except that a manual rather than a computer-programmable phase shifter is used because of its lower attenuation, and there is no non-invasive third port. 
    (b) For three selected perturbation strengths and the frequency interval highlighted by red bars in (a), the location of poles (+, blue) and zeros (x, red) 
    }
    \label{fig3}
\end{figure*}

Our starting point is the setup with hidden symmetry from Fig.~\ref{fig2}(c) and its symmetry-preserving tuning (we do not need the symmetry-breaking tuning nor the non-invasive third port in this section). Since the computer-controlled phase shifter in Fig.~\ref{fig2}(c) strongly attenuates the waves, we replace it with a manual mechanical phase shifter in this section whose attenuation is comparable to that of the coaxial cables. We now continuously tune the perturbation strength and measure the corresponding $S_{11}$ spectrum for each perturbation strength, yielding the perturbation-frequency map displayed in Fig.~\ref{fig3}(a). Despite the limited range of experimentally accessible perturbation strengths, it is apparent that, as expected, the observed patterns in Fig.~\ref{fig3}(a) repeat as the frequency increases. For instance, the pattern from the frequency interval around 980~MHz highlighted by red bars that we will focus on in the following is seen again in the vicinity of 1320~MHz. The selected interval displays two instances in which two troughs of the $|S_{11}|$ map cross at specific values of perturbation strength, hinting at the (approximate) existence of reflectionless EPs.

\setcounter{figure}{2}    
\begin{figure*}[h]
    \centering
    \includegraphics[width=0.0001\columnwidth]{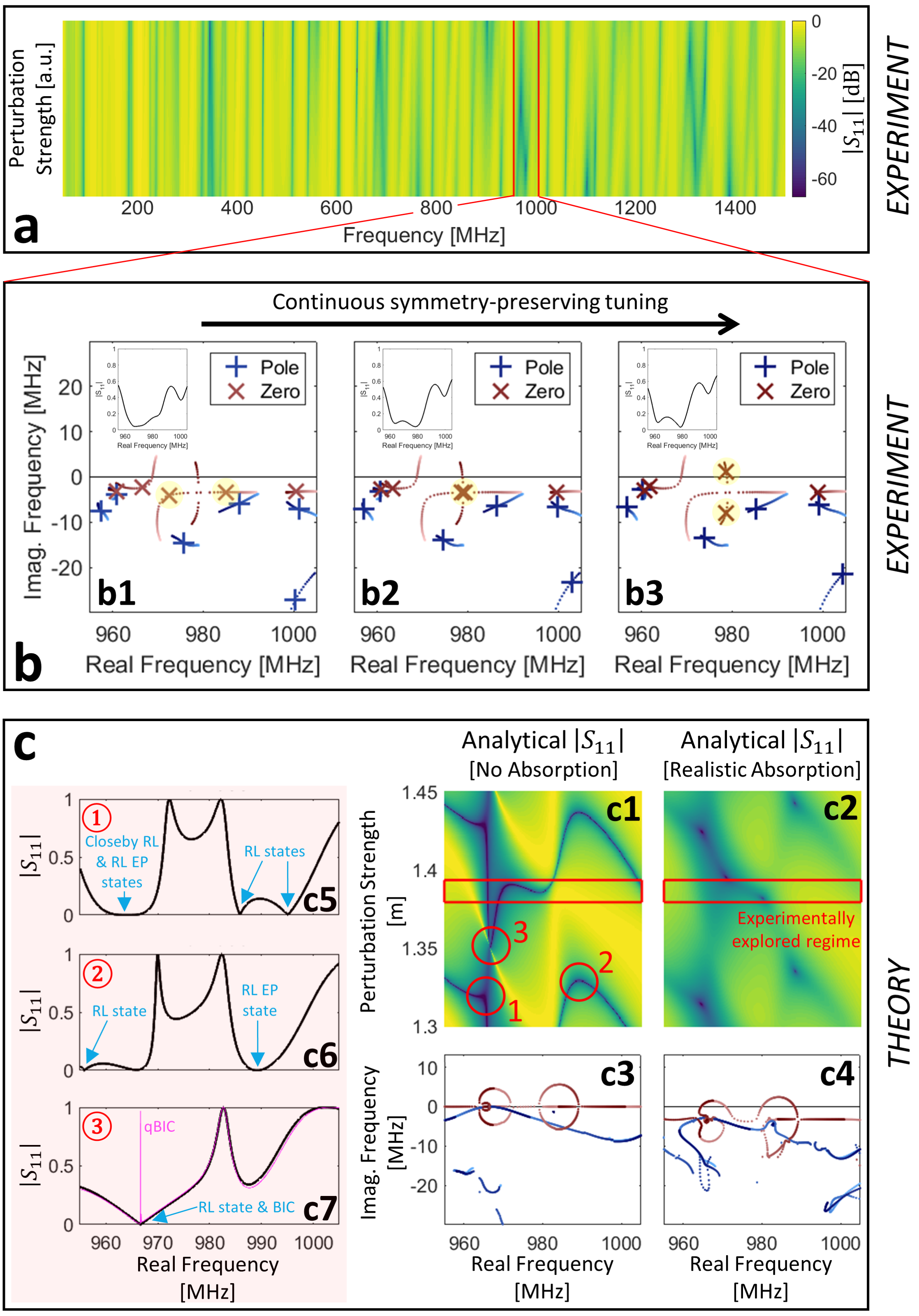}
    \caption{[CONTINUED] of $S_{11}$ are shown in the complex frequency plane. The two zeros of interest are highlighted in yellow. The poles' and zeros' trajectories upon continuous tuning are dotted and the color coding indicates the perturbation strength (light=low, dark=high). 
    $\mathcal{PT}$ symmetry would constrain the zeros to move along the real frequency axis upon tuning until they meet (giving rise to a reflectionless EP) and subsequently become a complex-conjugate pair of zeros. In our displayed experimental observation of these three cases, everything is shifted down in the complex frequency plane due to the finite amount of absorption.  
    (c) Analytical calculations of the perturbation-frequency map of $|S_{11}|$ over a wider range of perturbation strengths for the same setup without (c1) and with realistic (c2) absorption. The corresponding motion of poles and zeros is shown in (c3) and (c4) using the same color codes as in (b). For three selected perturbation strengths of particular interest, the corresponding $|S_{11}|$ spectra for the case without absorption are shown in (c5,c6,c7), featuring various reflectionless (RL) states and RL EP states. Moreover, in (c7) the system features a bound state in the continuum (BIC) which does not have any scattering signature; hence, the spectrum is also plotted for a slightly different perturbation strength, clearly showing the corresponding quasi-BIC (qBIC).}
    \label{fig3cont}
\end{figure*}

To confirm the existence of reflectionless EPs, we extract the location of poles and zeros of $S_{11}$ in the complex frequency plane~\footnote{Following standard practice in electronics and system control, we factorize our continuous-time transfer function $S_{11}$ in order to cast it as the ratio of two polynomials of the complex frequency variable~\cite{sanathanan1963transfer}. While this pole-zero factorization yields accurate results for our continuous-time analysis, it remains an approximation and for other wave engineering problems, in particular optical scattering problems, oftentimes more general formulations are needed to accurately describe the considered systems~\cite{grigoriev2013optimization,krasnok2019anomalies,zhang2020quasinormal,soltane2023multiple}.}. Poles and zeros correspond, respectively, to solutions of the wave equation with purely outgoing and incoming boundary conditions for the channel of interest (and purely outgoing boundary conditions for the remaining other channel). In passive systems without gain, poles are constrained to the lower half plane; the locations of zeros have no such constraint, and when a zero lies on the real frequency axis, reflectionless steady-state excitation of the system is possible at the corresponding real frequency. Under exact $\mathcal{PT}$ symmetry, the zeros of $S_{11}$ would be constrained to always lie on the real frequency axis, unless upon $\mathcal{PT}$-symmetry-preserving tuning a pair of zeros met on the real frequency axis (forming a reflectionless EP~\footnote{Perfectly absorbing EPs~\cite{sweeney2019perfectly,sweeney2020theory,wang2021coherent} also result from the coalescence of two zeros on the real frequency axis, but in situations involving excitation through all attached asymptotic scattering channels of a system with a finite amount of irreversible absorption.}) and upon further tuning left the real frequency axis as complex-conjugate pair~\cite{sweeney2020theory,ferise2022exceptional}. Although absorption breaks exact $\mathcal{PT}$ symmetry in our experiment, the level of absorption is sufficiently small such that the results from the case with exact $\mathcal{PT}$ symmetry can still be approximately observed.
In Fig.~\ref{fig3}(b1), we observe that the zeros of interest (highlighted in yellow) occur at complex frequencies with a negative imaginary part, i.e., they are firmly (but not strictly) constrained to moving along a horizontal line below the real frequency axis. Upon symmetry-preserving tuning, the two zeros almost meet on this horizontal line in Fig.~\ref{fig3}(b2). However, since they are not strictly confined to this line, they avoid crossing, and subsequently move (almost) vertically away from the line in opposite directions, eventually becoming an almost conjugate pair in Fig.~\ref{fig3}(b3). In the limit of zero absorption, the avoided crossing becomes the reflectionless EP described above and happens on the real frequency axis, making it accessible under steady-state excitation. Using the hidden $\mathcal{PT}$ symmetry, it is hence possible to conceive wave devices with functionalities such as broadened near-reflectionless excitation and higher-order signal differentiation without apparent mirror symmetry, endowing such wave devices with resilience against reverse-engineering threats.

Since we study a complex scattering system, as opposed to a simple Fabry-Perot-like system, other scattering singularities (zeros and poles) are present in the vicinity of our zeros of interest, as clearly seen in Fig.~\ref{fig3}(b). Since our zeros of interest are not strictly constrained to the horizontal line below the real frequency axis, interactions with other scattering singularities can slightly displace them from said horizontal line. 
Moreover, the lineshape of $|S_{11}|$ cannot be clearly analyzed due to the influence of other scattering singularities in close proximity, as seen in the insets in Fig.~\ref{fig3}(c). 
Given the finite amount of absorption, the lowest reflection is observed in the so-called $\mathcal{PT}$-broken phase when the upper zero crosses the real frequency axis~\footnote{Incidentally, the presence of a zero in the upper plane, resulting in a so-called discontinuity branch bridging the zero and a pole across the real frequency axis, has recently been shown to shed new light on the design of Huygens metasurfaces~\cite{colom2022crossing}.}.

To further explore the implemented covert hidden-$\mathcal{PT}$-symmetry-based scattering control, we now analytically calculate based on Eqs.~(\ref{eq_s_matrix_gen},\ref{eq_quantumgraphsmatrix}) the spectra of $S_{11}$ for a wider range of perturbation strengths, once for the case with exact $\mathcal{PT}$ symmetry (no absorption, Fig.~\ref{fig3}(c1)) and once with the frequency-dependent absorption extracted from our experiments (Fig.~\ref{fig3}(c2)). We observe, as expected, that the perturbation-frequency maps repeat not only along the frequency axis but are also periodic along the perturbation strength axis. Multiple reflectionless EPs are apparent in Fig.~\ref{fig3}(c1) and the corresponding singularity plot in Fig.~\ref{fig3}(c3): one is seen to occur within the experimentally explored regime highlighted by the red box, as observed in Fig.~\ref{fig3}(a,b), and three additional ones are highlighted by red circles. Of these, the one indexed \textcircled{2} is another ``regular'' reflectionless EP and displays the expected quadratic U-shaped $|S_{11}|$ dip on the real frequency axis in Fig.~\ref{fig3}(c6). This is exactly the transfer function that is needed to implement second-order differentiation of a time-domain signal  whereas the V-shaped dips also seen in Fig.~\ref{fig3}(c) (corresponding to a single real-valued zero) enable first-order temporal differentiation~\cite{sol2022meta,ferise2022exceptional}.
In contrast, the reflectionless EP indexed \textcircled{1} occurs in close proximity to another reflectionless state, giving rise to an unusually wide reflection dip in Fig.~\ref{fig3}(c5).

The most interesting case, however, is that indexed \textcircled{3}. Here, two zeros and a pole meet at the same location on the real frequency axis. Recall that poles are not allowed on or above the real frequency axis in passive system -- unless they are part of a bound state in the continuum (BIC)~\cite{hsu2016bound}. BICs, also known as trapped modes, contain neither incoming nor outgoing radiation and exist when a zero and a pole coincide on the real frequency axis, implying that their topological charges annihilate~\cite{sweeney2020theory,sakotic2023non}. 
Indeed, the $|S_{11}|$ spectrum in Fig.~\ref{fig3}(c7) displays a V-shaped dip typical for a single real-valued zero because BICs cannot be seen in scattering spectra as they do not couple to the asymptotic scattering channels. However, upon a slight perturbation, the scattering spectrum will feature an ultrathin peak, as seen on the purple $|S_{11}|$ spectrum in Fig.~\ref{fig3}(c7). The sharp peak of a qBIC, mainly studied in the context of flat-optics metasurfaces to date~\cite{koshelev2018asymmetric}, has enticing potential applications ranging from sensing to signal filtering. The presence of a BIC in our system with \textit{hidden} mirror symmetry comes as no surprise since mirror-symmetric systems are known to feature BICs~\cite{evans1994existence,dhia2018trapped}. Our covert hidden-$\mathcal{PT}$-symmetry-based scattering control enables the conception of devices featuring qBICs without apparent symmetry, yielding resilience to reverse engineering threats.

\section{Meta-Atoms with Non-Reciprocal Interactions}
\label{sec_nonreciprocal}

Before closing, we ask in this section whether the covert scattering control demonstrated in the previous sections is limited to metamaterials in which the direct interactions between meta-atoms are reciprocal. The inclusion of nonreciprocal interactions between meta-atoms, e.g., due to isolators or circulators, is common in radiofrequency and nanophotonic networks, for instance, to prevent reflected-power echoes. Quantum graphs involving circulators have also been explored to conceive topology-protected wave devices~\cite{zhang2021superior} and as experimental platforms to study the statistical properties of certain classes of quantum chaos~\cite{lawniczak2010experimental,bialous2016power,rehemanjiang2016microwave,lawniczak2019investigation,chen2021generalization,zhang2022experimental}; moreover, Ref.~\cite{chen2020perfect} showed that a quantum graph including a circulator can be tuned to have a real-valued zero. However, existing theoretical works on wave scattering control regarding hidden symmetries or regarding reflectionless EPs do not cover the case of non-reciprocal interactions between meta-atoms. 
In this section, by providing an example of a cable-network metamaterial with hidden $\mathcal{P}$ symmetry despite the inclusion of circulators, we demonstrate that our previous results are \textit{not} limited to purely reciprocal constituents in the metamaterial. 
We furthermore analyze an example of a non-reciprocal cable-network metamaterial (without hidden $\mathcal{P}$ symmetry) that is both equi-reflectional and $\mathcal{PT}$ symmetric.

If the magnetic vector potential $A_{i,j} = -A_{j,i}$ acting on the bond (interaction) between two meta-atoms indexed $i$ and $j$ is non-zero, the phase of the transmission across this bond will be direction-dependent and hence non-reciprocal -- see Eq.~(\ref{eq_quantumgraphsmatrix}). Conventional microwave circulators like the ones we use below isolate certain ports from others by creating destructive interferences based on this principle. In the analytical framework, a circulator is consequently treated as three auxiliary meta-atoms with non-reciprocal interactions, as detailed in Supplementary Note V. 
Such circulators cannot display an ``ideal behavior'' over an extended frequency range, as we discuss in more detail in Supplementary Note V. Specifically, they will feature an imperfect isolation and a non-zero delay.

\begin{figure*}[htbp]
    \centering
    \includegraphics[width=\columnwidth]{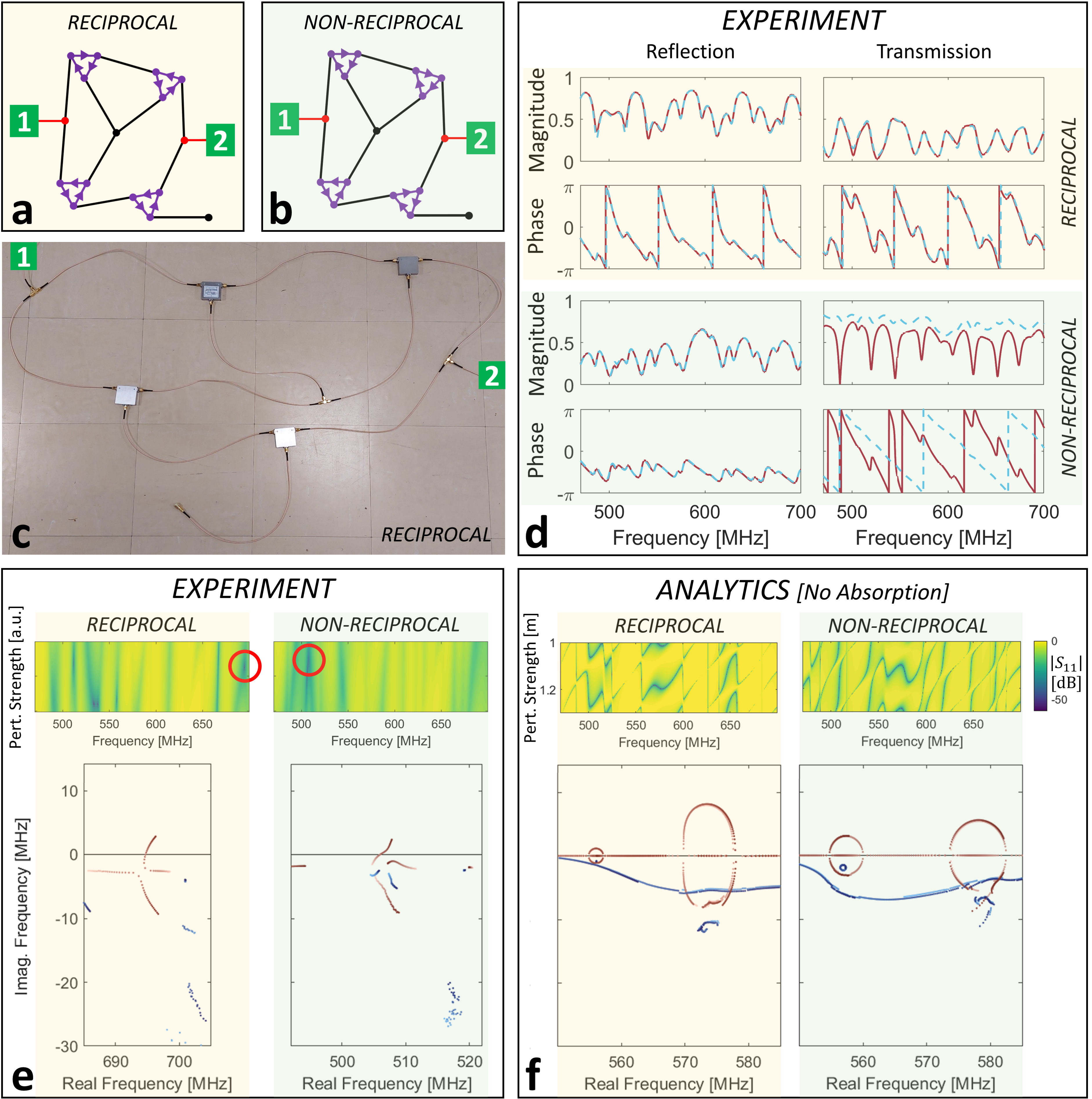}
    \caption{\textbf{Equi-Reflectionality and Reflectionless Exceptional Points in Metamaterials with Non-Reciprocal Components.} 
    (a,b) Topology of equi-reflectional \textit{reciprocal} (a) and \textit{non-reciprocal}  cable-network metamaterials involving circulators.
    (c) Photographic image of the experimental implementation of the metamaterial from (a).
    (d) Scattering measurements of the two metamaterials from (a,b). The overlayed scattering coefficients visualize equi-reflectionality and (non-)reciprocity.
    (e) Experimental observation of reflectionless EPs in the reciprocal and non-reciprocal metamaterials involving circulators.
    (f) Analytical scattering calculations (zero absorption) for the two metamaterials involving circulator-like devices. Multiple reflectionless EPs are seen in each case.}
    \label{fig4}
\end{figure*}

The metamaterials explored in previous sections were all naturally reciprocal (regarding their transmission spectra) since they contained nothing that could break reciprocity. If, however, the metamaterial contains circulators, its scattering matrix is neither guaranteed to be reciprocal nor is it guaranteed to be non-reciprocal. Both options are in principle conceivable for two-port cable-network metamaterials including circulators that feature the equi-reflection property: 

\textit{1) Reciprocal} ($S_{21}=S_{12}$). If the metamaterial is reciprocal despite the inclusion of circulators, there is no way to determine from outside the system that the system includes non-reciprocal components. Specifically, the system's scattering properties can be fully captured by a reduced-basis representation without any non-reciprocal components. Consequently, the system must behave exactly like the metamaterials without circulators studied in the previous sections, enabling in particular covert access to reflectionless EPs via continuous tuning of a single symmetry-preserving parameter (in the case of zero absorption). 

\textit{2) Non-Reciprocal} ($S_{21} \neq S_{12}$). If the metamaterial is non-reciprocal, it cannot have $\mathcal{P}$ symmetry in the reduced basis. However, $\mathcal{P}$ symmetry is not a necessary condition for equi-reflectionality, as we will discuss below, so that non-reciprocity of the transmission does not exclude the possibility of equi-reflectionality. Moreover, assuming zero absorption and the absence of gain mechanisms, the lack of $\mathcal{P}$ symmetry does not exclude the possibility of a $\mathcal{PT}$ symmetry that would enable the covert access to reflectionless EPs. Note that under the assumption of zero absorption, the non-reciprocity of a two-channel scattering matrix can only be phase-wise~\cite{caloz2018electromagnetic}.

A brute-force search of two-port cable-network metamaterials involving circulators that are equi-reflectional irrespective of the amount of isolation and the propagation delay between circulator ports (but assuming isolation and delay do not differ between different pairs of circulator ports, and that all circulators are identical) yielded the two structures shown in Fig.~\ref{fig4}(a,b). Strikingly, they only differ in the orientation of two circulators, and one of these two metamaterials is reciprocal whereas the other one is non-reciprocal. The existence of these examples proves that both possible above-described classes of metamaterials involving circulators that feature hidden equi-reflectionality hence exist. 
Our experimental measurements shown in Fig.~\ref{fig4}(d) validate this fact. The transmissions in the non-reciprocal case are seen to differ in terms of phase and magnitude which is allowed in our system given its finite amount of absorption. In fact, we have observed that the transmission time delay in one direction is on average roughly twice the transmission time delay in the other direction, so that the larger dwell time and hence exposure to inevitable absorption plausibly explains why the transmission amplitude is much lower in the former case. 
We detail in Supplementary Note~II.G the limited isolation and finite propagation delay of the circulators used in our experiments.

The non-reciprocal case evidences that for 2-port systems the broadband equi-reflection property is in fact possible without any apparent or hidden $\mathcal{P}$ symmetry. The inverse of any $2 \times 2$ matrix with equal diagonal entries will also have equal diagonal entries. Hence, if the reduced-basis matrix $\mathbf{\tilde{H}}=\mathbf{\tilde{A}_0}-\mathbf{\Delta_{\mathcal{S}\mathcal{S}}}
$ has equal diagonal entries ($\tilde{H}_{1,1} = \tilde{H}_{2,2}$), and each of the two asymptotic scattering channels is coupled to one of the primary meta-atoms, then it follows that $S_{1,1} = S_{2,2}$, irrespective of the off-diagonal entries of $\mathbf{\tilde{H}}$. This is precisely the case for the example from Fig.~\ref{fig4}(b). Note that the independence of the diagonal entries of $\mathbf{S}$ from the off-diagonal entries of $\mathbf{\tilde{H}}$ seen here is special to the case of $m=2$ primary meta-atoms, and not in general valid for $m>2$. Moreover, based on analytical scattering calculations, we have confirmed that there is nonetheless a hidden $\mathcal{PT}$ symmetry at all frequencies in $\mathbf{\tilde{H}}$. Note that $\mathcal{PT}$ symmetry alone is in general not associated with equi-reflectionality.

Since both the reciprocal and the non-reciprocal case feature a hidden $\mathcal{PT}$ symmetry (under the assumption of no absorption), we expect to be able to observe reflectionless EPs in both cases upon symmetry-preserving tuning. Reflectionless EPs have not been previously studied in either class of such scattering systems involving non-reciprocal components, irrespective of whether the $\mathcal{PT}$ symmetry is apparent or hidden. 
Using the same symmetry-preserving tuning mechanism as in Fig.~\ref{fig3}, we show in Fig.~\ref{fig4}(f) that indeed we observe reflectionless EPs in both cases in analytical scattering calculations assuming a plausible circulator-like device (see Supplementary Note V) and no absorption. (Our analytical scattering calculations cannot directly match our experiments here because of the unknown circulator parameters.)  
In our experiments displayed in Fig.~\ref{fig4}(e), we also approximately observe the reflectionless EPs, within the same limitations due to non-zero absorption as in the previous section. In the case of the non-reciprocal metamaterial, the presence of two poles and the arrival of a third zero in the vicinity of the meeting point of the two zeros of interest somewhat distorts their encounter from the usual pattern (which is possible since they are not firmly constrained to a horizontal line due to the non-zero absorption), but the approximate property of the two zeros taking off from a horizontal line in opposite vertical directions after their encounter is still apparent.

Finally, we mention a peculiarity that we observe in our analytical scattering calculations for the non-reciprocal equi-reflectional metamaterial (see Supplementary Note VI for details). Under the assumption of ideal broadband circulators, which seems to describe an unrealistic physical system (see Supplementary Note V), the specific symmetry-preserving tuning mechanism used throughout this paper appears to never yield a reflectionless EP because of an apparent repulsion between zeros. This qualitatively new behavior is intriguing but presumably not relevant to the experimental reality.

\section{Conclusion}

To summarize, we have shown that wave devices and systems whose functionality is based on a combination of $\mathcal{P}$ or $\mathcal{PT}$ symmetry with tunability can be implemented without apparent symmetry, making them resilient against reverse-engineering threats. The key conceptual insight underlying this covert wave scattering control is that ultimately the (local and non-local) coupling between the primary meta-atoms (those meta-atoms that are directly coupled to asymptotic scattering channels) matters; a $\mathcal{P}$ symmetry in the reduced basis of primary meta-atoms can be absent in the canonical basis (and hence hidden from sight) through a suitably chosen topology of non-local interactions between the primary meta-atoms mediated by the secondary meta-atoms. We have applied this covert scattering control in microwave experiments with cable-network metamaterials to demonstrate a scheme for physical-layer secure communications as well as the conception of reverse-engineering-resilient wave devices featuring reflectionless exceptional points and quasi bound states in the continuum. Moreover, we discovered that metamaterials with hidden symmetries can also include non-reciprocal components like circulators and, for the first time, we observed reflectionless exceptional points in a non-reciprocal system.

Looking forward, given the generality of the underlying wave concepts, our work can be transposed to different wave platforms spanning from acoustics and mechanics, via photonics and electronics, all the way to quantum mechanics. Of particular interest are technologically relevant fully integrated electronic systems whose ability to compensate the inevitable absorption has already been demonstrated. 
Moreover, wavefront shaping can yield new degrees of control over the coupling of the asymptotic scattering channels to the primary meta-atoms in (nano)photonic systems.

\begin{acknowledgments}
\vspace{-0.5cm}

M.R. and P.d.H. acknowledge stimulating discussions with V.~Pagneux. P.d.H. furthermore acknowledges stimulating discussions with P.~Genevet and A.~D.~Stone.

P.d.H. acknowledges funding from the CNRS pr\'{e}maturation program (project ``MetaFilt''), the ANR PRCI program (project ANR-22-CE93-0010-01), the European Union's European Regional Development Fund, and the French region of Brittany and Rennes Métropole through the contrats de plan État-Région program (project ``SOPHIE/STIC \& Ondes'').

\end{acknowledgments}

\vspace{-0.5cm}

\section*{Conflict of Interest}
\vspace{-0.5cm}

The authors declare no conflict of interest.

\vspace{-0.5cm}

\section*{Author Contributions}
\vspace{-0.5cm}

P.d.H. conceived the project. P.d.H. developed the generalized theory of scattering systems with hidden symmetries and conducted the analytical scattering calculations. J.S. and P.d.H. conducted the experiments. M.R. conducted the search of topologies of cable-network metamaterials involving circulators that feature the equi-reflection property in Sec.~\ref{sec_nonreciprocal}. P.d.H. performed the data analysis and wrote the manuscript. All authors contributed with thorough discussions.

\clearpage

\providecommand{\noopsort}[1]{}\providecommand{\singleletter}[1]{#1}%

\end{document}